\documentclass[journal=apchd5,manuscript=article]{achemso}

\usepackage[T1]{fontenc}

\usepackage{geometry}
\usepackage{setspace}

\usepackage{xcolor}
\usepackage{amssymb}

\usepackage{graphicx}
\usepackage{float}

\newfloat{scheme}{htbp}{los}
\floatname{scheme}{Scheme}
\floatname{chart}{Chart}
\newfloat{graph}{htbp}{loh}

\usepackage[version = 4]{mhchem} 

\author{Sachin Sharm}
\affiliation{Department of Physics \& Astronomy, Texas Tech University, Lubbock, TX 79409, USA}

\author{Elliott Walker}
\affiliation{Department of Physics \& Astronomy, Texas Tech University, Lubbock, TX 79409, USA.; Department of Mathematics, Texas Tech University, Lubbock, TX 79409, USA}

\author{Rachael Myers-Ward}
\affiliation{U.S. Naval Research Laboratory, Washington, D.C. 20375, USA}

\author{Jenifer Hajzus}
\affiliation{U.S. Naval Research Laboratory, Washington, D.C. 20375, USA}

\author{Yijing Liu}
\affiliation{Georgetown University, Washington, D.C. 20057, USA}

\author{Paola Barbara}
\affiliation{Georgetown University, Washington, D.C. 20057, USA}

\author{Ioannis Chatzakis}
\email{*ioannis.chatzakis@ttu.edu}
\affiliation{Department of Physics \& Astronomy, Texas Tech University, Lubbock, TX 79409, USA}

\title{Excitation-Energy-Selective Control of Hot-Carrier Cooling via a Resonant Optical-Phonon Bottleneck in Graphene}

\keywords{electron-phonon, hot-carriers cooling, phonon bottleneck, phonon dynamics}

\begin{document}

\maketitle

\begin{abstract}

Understanding and controlling hot-carrier relaxation in graphene is crucial for advancing ultrafast optoelectronic and terahertz technologies. Here, we investigate carrier cooling dynamics in mono- and bilayer graphene using mid-infrared pump pulses (0.22 to 0.73 eV) and terahertz probe pulses. We uncover a pronounced, reproducible, and non-monotonic dependence of the carrier relaxation time on excitation photon energy. Remarkably, within a narrow spectral window (0.42 to 0.48 eV), the carrier lifetime increases by an order of magnitude compared to a few picosecond-scale cooling observed at other energies. We show that this anomalous slowdown originates from a resonant enhancement of the optical-phonon lifetime, causing accumulation and reabsorption of hot optical phonons that suppress energy transfer to the lattice. All observed behaviors are captured within a unified carrier-phonon energy-balance framework, where excitation-energy-dependent variations of the effective optical-phonon decay pathway govern the cooling dynamics. These findings demonstrate excitation-energy-selective control of hot-carrier relaxation in graphene and provide new insight into non-equilibrium carrier-phonon interactions near the optical-phonon bottleneck.

\end{abstract}

\vspace{1cm}

Understanding how non-equilibrium carriers relax in solids is central to the operation of ultrafast electronic, optoelectronic, and terahertz devices. In low-dimensional materials, reduced phase space and strong many-body interactions profoundly modify carrier-phonon coupling, enabling relaxation pathways that differ qualitatively from those in conventional semiconductors and metals  \cite{bib1, bib2, bib3}. Graphene, with its linear band structure, high carrier mobility, and strong coupling to optical phonons, has therefore emerged as a prototypical platform for exploring hot-carrier dynamics far from equilibrium   \cite{bib4, bib5, bib6}.
Following optical excitation, photoexcited carriers in graphene rapidly thermalize through carrier-carrier scattering, establishing a hot Fermi-Dirac distribution on sub-100-fs timescales\cite{bib7,bib8,bib9}. Subsequent energy relaxation proceeds primarily through emission of strongly coupled optical phonons near the $\Gamma$ and $K$ points, which then decay into acoustic phonons and the substrate \cite{bib10,bib11,bib12,bib13}. This multistep cooling process has been extensively investigated using ultrafast optical and terahertz spectroscopy, revealing relaxation times ranging from hundreds of femtoseconds to several picoseconds depending on carrier density, excitation energy, and temperature\cite{bib14,bib15,bib16, Tielrooij, Sharma, Shi, Tomadin, Jnawali, Frenzel, Hale, bib17,bib18}. A long-standing challenge, however, has been to identify practical control parameters that allow carrier cooling to be reversibly tuned, rather than merely modified, under ambient conditions.
Previous studies have demonstrated that carrier relaxation in graphene depends sensitively on the Fermi energy, reflecting the competition between interband and intraband excitation pathways and the influence of Pauli blocking\cite{bib19,bib20,bib21,bib22}. Time-resolved terahertz photoconductivity measurements have revealed both positive and negative photoresponses, highlighting the balance between carrier heating and enhanced phonon scattering\cite{Tielrooij,bib24,bib25,bib26}. Theoretical work has reconciled these observations by showing that interband transitions dominate at low doping, while intraband heating governs the dynamics at higher Fermi energies\cite{bib27,bib28}. Experiments employing a broad range of pump photon energies from the near-infrared to the terahertz have further emphasized the importance of Coulomb and electron-phonon interactions, and have reported signatures of suppressed relaxation when the excitation energy approaches the optical-phonon threshold\cite{bib30,bib31,bib32}. Despite these advances, it remains unclear whether hot-phonon effects can be selectively activated and controlled by excitation energy alone at room temperature.
Here we report a pronounced and highly energy-selective suppression of hot-carrier cooling in graphene, observed using mid-infrared pump-terahertz probe spectroscopy. By tuning the excitation photon energy over a wide range while independently controlling carrier density and pump fluence, we identify a narrow excitation window (0.42 - 0.48 eV) in which carrier relaxation times increase dramatically, exceeding those measured at both lower and higher photon energies by severalfold. Outside this window, cooling remains fast and exhibits only weak or saturating fluence dependence, even at comparable excitation densities.
We show that this anomalous slowdown is population-driven and arises from the buildup and reabsorption of a non-equilibrium optical-phonon population, forming a hot-phonon bottleneck that is selectively activated near the center of the excitation window \cite{bib11,bib12,bib33}. Crucially, all observed regimes are captured within a single extended carrier-phonon energy-balance model \cite{Allen,Perfetti,Ishida} in which the governing equations remain unchanged, while the effective optical-phonon lifetime varies with excitation photon energy. Our results establish excitation energy as a powerful and previously unexplored control knob for hot-carrier relaxation in graphene and highlight the central role of optical-phonon dynamics in engineering ultrafast energy flow in two-dimensional materials.


Here we investigated hot-carrier relaxation in monolayer and bilayer graphene using mid-infrared pump pulses with photon energies spanning 0.37 - 0.72~eV and terahertz time-domain spectroscopy (THz-TDS) as a probe. The principle of the experimental configuration is illustrated in Fig. 1, together with the measured terahertz (THz) transient in the time domain and the corresponding frequency spectrum. Following optical excitation, the pump-induced change in THz transmission exhibits a prompt response followed by a decay on picosecond timescales. For all excitation energies, the transient signals are well described by single-exponential decays at delays exceeding the initial electronic thermalization time, allowing extraction of a characteristic relaxation time ($\tau$).

\begin{figure}[!htbp]
\centering
\includegraphics[scale=0.6]{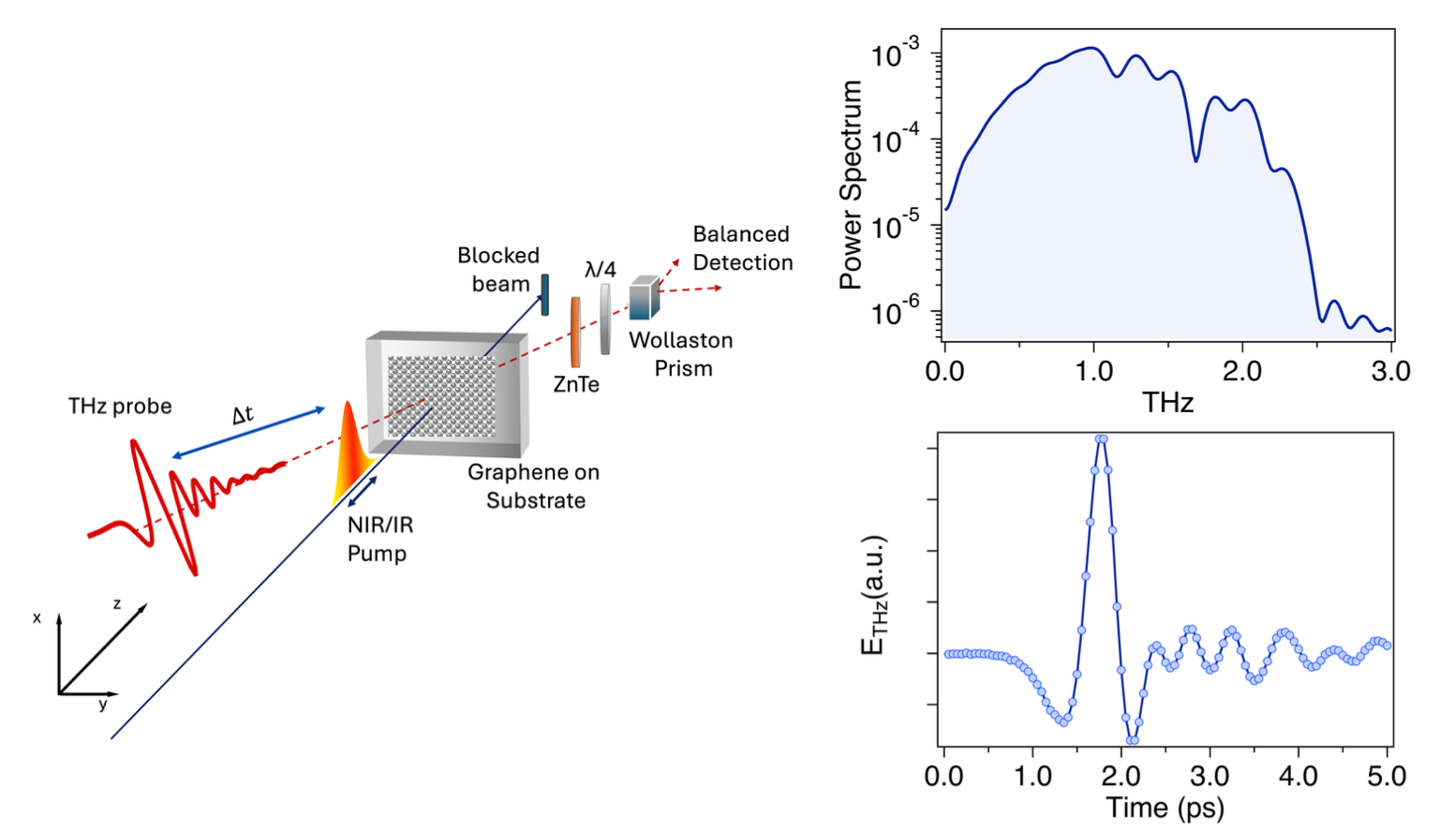}
\caption{(Left) Schematic of the THz pump-probe setup and the components for THz generation and detection. (Right) Time-domain THz waveform (bottom) and corresponding Fourier-transform spectrum (top), extending to 2.5 THz.}\label{Pumpprobe}
\end{figure}

The experiments were conducted on monolayer and bilayer graphene samples prepared using two complementary growth techniques. Monolayer graphene samples were prepared by chemical vapor deposition (CVD) and transferred onto SiO$_2$/Si substrates with a 280 nm oxide layer. Bilayer graphene samples were quasi-freestanding bilayer graphene grown epitaxially on semi-insulating 6H-SiC(0001) substrates by thermal decomposition, followed by hydrogen intercalation to decouple the graphene from the substrate. Details can be found in the supporting information.

For one monolayer sample, the carrier density was determined by Hall measurements to be $2.2\times10^{11}$~cm$^{-2}$, corresponding to a Fermi energy of approximately 0.06~eV. For a second monolayer sample, the Fermi energy was estimated to be approximately 0.18~eV from Raman spectroscopy. All samples were measured under ambient conditions at room temperature. 

Two bilayer epitaxial graphene samples were investigated, with sheet carrier densities determined by Hall measurements to be $7.9 \times 10^{12}$~cm$^{-2}$ and $9.1 \times 10^{12}$~cm$^{-2}$, corresponding to Fermi energies of approximately 0.36~eV and 0.38~eV, respectively. The corresponding room-temperature mobilities were 4290~cm$^2$V$^{-1}$s$^{-1}$ and 1650~cm$^2$V$^{-1}$s$^{-1}$.

Ultrafast measurements were performed using a typical optical pump-THz probe setup driven by a Ti:sapphire regenerative amplifier operating at a repetition rate of 1 kHz, combined with an optical parametric amplifier to generate the mid-IR pump pulses. Details can be found in the supporting information file.  
In addition, control experiments were performed on bare SiC, sapphire, and SiO2/Si substrates under identical excitation conditions. No pump-induced   THz signal was detected in these measurements, confirming that all the observed dynamics originate from the graphene layers.

To establish the generality of the excitation energy dependence of the relaxation of hot carriers , we systematically measured the relaxation time as a function of excitation photon energy for multiple graphene samples with different layer numbers, substrates, and Fermi energies. The extracted relaxation times are summarized in Figure 2 (a). Across all samples, $\tau$ exhibits a pronounced non-monotonic dependence on excitation energy. For excitation energies below approximately 0.42 eV and above $\sim$~ 0.48eV, the relaxation time remains confined to the range of $\sim$1-4 ~ps (Fig. 2b). In contrast, a sharp enhancement of $\tau$ is observed within a narrow window between $\sim$ 0.42 and $\sim$ 0.48 eV, with the longest lifetimes consistently occurring near $\sim$ 0.45 eV.

\begin{figure}[!htbp]
\centering
\includegraphics[scale=0.45]{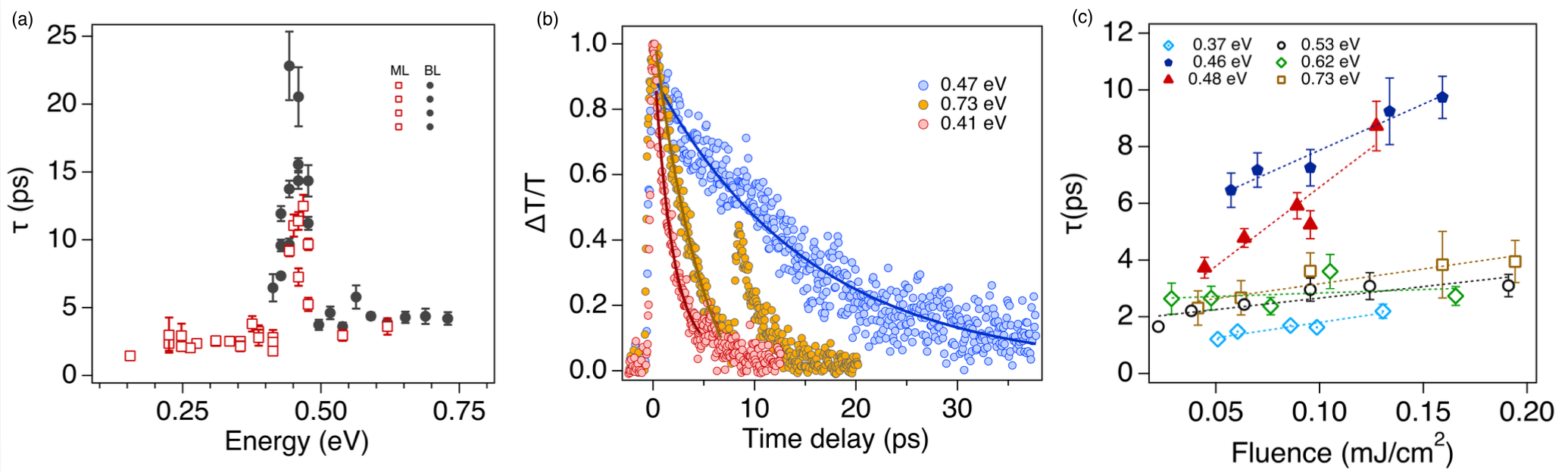}
\caption{(a) Relaxation time of photoexcited carriers in graphene as a function of excitation photon energy, extracted from pump-probe THz measurements. The relaxation time exhibits a pronounced non-monotonic dependence on photon energy. Energy-interval averaging yields mean lifetimes of $\langle \tau \rangle \approx 3.2$~ps for $E < 0.42$~eV and $\langle \tau \rangle \approx 3.8$~ps for $E > 0.48$~eV, indicating predominantly fast relaxation outside the mid-energy region. In contrast, an anomalous energy-selective slowing of carrier relaxation emerges within a narrow spectral window between $\approx 0.42$ and $0.48$~eV, where lifetimes increase to $\tau \approx 10$-$25$~ps ($\langle \tau \rangle \approx 18$~ps), reaching values up to $\approx 20$~ps - nearly an order of magnitude longer than the few-picosecond decay observed at lower and higher excitation energies. (b) Time-resolved differential transmission of graphene following ultrafast optical excitation. The experimentally measured time-domain $\Delta T(t)/T$ is shown for three representative pump photon energies (0.41, 0.47, 073 eV), below, within, and above the resonance window centered at approximately 0.45 eV. The solid lines represent fits to a single-exponential decay function. The extracted decay constants define the carrier relaxation times and reveal a strong dependence on excitation photon energy, with anomalously long relaxation observed when the excitation energy lies within the resonant window near 0.45 eV. Secondary peaks appearing at approximately 5.8 ps (red trace) and 9 ps (yellow trace) arise from the round trip of the THz probe pulse within the substrate, which subsequently re-excites the graphene layer. (c) The relaxation time is plotted as a function of excitation fluence for several excitation photon energies. The dashed lines represent linear fits to the data. Notably, the slopes corresponding to excitation wavelengths in the resonant regime (filled data points) are significantly steeper, confirming that the excitation photon energy plays a key role in determining the relaxation dynamics.}\label{Fluencedependence}\label{Fuencedependence}
\end{figure}

Notably, this enhancement does not track monotonically with excitation energy. Instead, the relaxation time peaks within the intermediate window and decreases again at higher photon energies, demonstrating that the observed slowdown is confined to a specific excitation-energy range rather than arising from a gradual evolution of carrier cooling efficiency.

To further characterize the nature of the anomalous slowdown, we examined the dependence of the relaxation time on pump fluence at selected excitation energies. Figure 2 (c) shows $\tau$ as a function of fluence for excitation energies below ($\sim$ 0.37 and 0.41 eV), within ($\sim$ 0.45 and 0.48 eV), and above ($\sim$ 0.53, 0.62, and 0.72 eV) the intermediate window.

Outside the 0.42 - 0.48~eV range, the relaxation time exhibits only weak or saturating fluence dependence. For excitation at 0.37 and 0.41~eV, $\tau$ increases modestly with fluence and saturates in the 1--3 ~ps range. Similarly, for excitation at $\sim$ 0.53, 0.62, and 0.72~eV, $\tau$ remains in the few-picosecond regime, showing at most a moderate and gradual increase with fluence.

In contrast, excitation within the intermediate window produces a qualitatively different behavior. At $\sim$ 0.45 eV, the relaxation time increases strongly and monotonically with fluence, reaching values several times larger than those observed outside the window. Excitation at $\sim$ 0.48 eV yields intermediate behavior, with relaxation times longer than those at neighboring energies but shorter than those observed at $\sim$ 0.45 eV. This comparison indicates that the anomalous slowdown is maximized near $\sim$ 0.45~eV and weakens toward the upper edge of the window.

To determine whether the observed energy-dependent relaxation originates from changes in optical absorption or probe-induced effects, we analyzed the pump-induced change in the THz conductivity. For all excitation energies and fluences investigated, the photoinduced conductivity is negative and exhibits no anomalous spectral features across the THz bandwidth. The magnitude of the conductivity change varies smoothly with excitation energy and fluence, with no enhancement coinciding with the window of prolonged relaxation.

In addition, control measurements performed on bare substrates show no pump-induced THz response, confirming that the observed dynamics arise from the graphene layers. The absence of any anomaly in the conductivity amplitude demonstrates that the enhanced relaxation time within the 0.42--0.48~eV window is not associated with changes in absorption, carrier density, or probe coupling, but instead reflects a modification of the relaxation dynamics.

Taken together, the results establish three robust experimental facts: (i) carrier cooling in graphene is fast and weakly fluence dependent over a broad range of excitation energies; (ii) a pronounced and reproducible slowdown emerges only within a narrow excitation window centered near$\sim$ 0.45 eV; and (iii) this slowdown is dynamical in nature, as it is not accompanied by anomalous changes in the photoinduced conductivity. These observations form the experimental basis for the subsequent interpretation.

The complex THz transmission coefficient $T^*(\omega)$ was obtained from the ratio of the Fourier-transformed THz electric fields transmitted through the graphene sample, $E_{\mathrm{gr}}(\omega)$, and through a reference substrate, $E_{\mathrm{ref}}(\omega)$. The frequency-dependent complex sheet conductivity $\sigma(\omega)$ was extracted using the thin-film approximation \cite{Tinkham}

 \begin{equation}
T^*(\omega) = \frac{E_{\mathrm{gr}}(\omega)}{E_{\mathrm{ref}}(\omega)}  
= \frac{1+n}{1+n + Z_0 \sigma(\omega)},
\end{equation}

where $n$ is the refractive index of the substrate ($n=3.124$ for SiC and $n=3.42$ for Si), and $Z_0 = 377~\Omega$ is the vacuum impedance. The contribution of the $280$~nm SiO$_2$ layer was neglected, as its thickness is much smaller than the THz wavelength.

Pump-induced changes in THz transmission were quantified as
\begin{equation}
\frac{\Delta T}{T} = \frac{T_{\mathrm{on}} - T_{\mathrm{off}}}{T_{\mathrm{off}}},
\end{equation}
where $T_{\mathrm{on}}$ and $T_{\mathrm{off}}$ are the transmitted THz signals with and without optical excitation, respectively. For thin conducting films, $\Delta T/T$ is directly proportional to the pump-induced change in sheet conductivity $\Delta\sigma$. A negative $\Delta\sigma$ was consistently observed for all samples and excitation energies, indicating intraband-dominated carrier heating. The measured differential transmission, $\Delta T / T_0$, is related to the photoinduced conductivity by
\begin{equation}
\Delta \sigma = \frac{1+n}{Z_0} \left( \frac{1}{1+\Delta T/T_0} - 1 \right).
\end{equation}
The experimental data were fitted using an exponential function.

\begin{figure}[!htbp]
\centering
\includegraphics[scale=0.6]{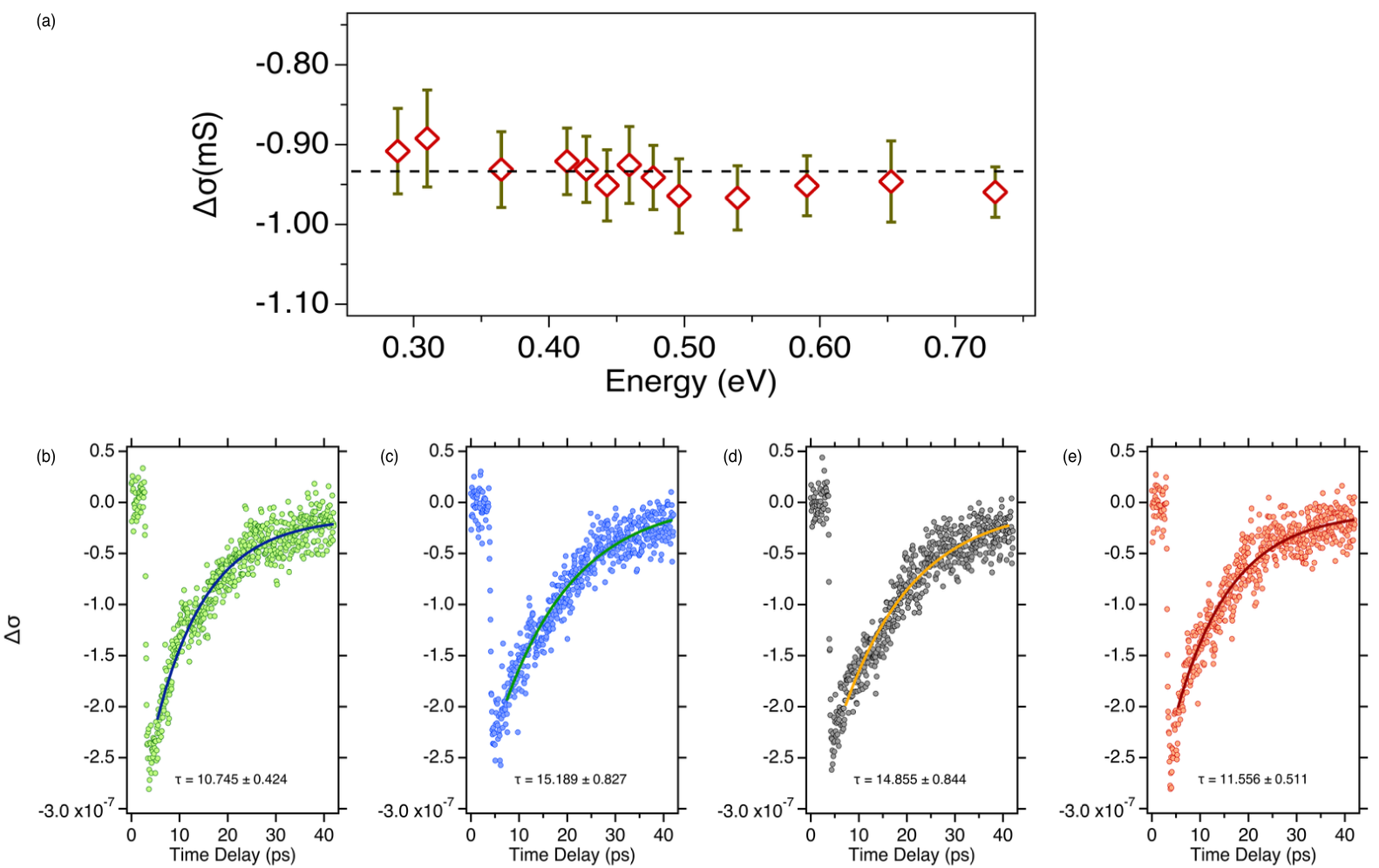}
\caption{(a) Energy-dependent photoinduced change in the THz conductivity, $\Delta \sigma$, of graphene following ultrafast optical excitation. Each data point represents the average of the frequency-domain conductivity extracted from individual THz spectra collected at the corresponding pump photon energy. The traces (b -- e) depict the dynamics of the photoinduced conductivity that was extracted from the measured $\Delta T/T $ for energies approximately 0.42, 0.44, 0.45, and 0.47 eV respectively.}\label{Fluencedependence}
\end{figure}

The ultrafast carrier dynamics are modeled using an extended two-temperature model that explicitly accounts for a non-equilibrium population of strongly coupled\cite{Allen,Perfetti,Ishida,Anisimov,Chatzakis} optical phonons. In this approach, the electronic system with temperature $T_e(t)$ exchanges energy with a hot optical-phonon bath characterized by an effective temperature $T_{\mathrm{op}}(t)$, while the remaining lattice degrees of freedom (acoustic phonons and substrate) are treated as a thermal reservoir at a fixed temperature $T_l$. The coupled energy-balance equations are
\begin{align}
C_e \frac{dT_e}{dt} &= S(t) - G_{e\text{-}op}\left(T_e - T_{\mathrm{op}}\right) - G_{\mathrm{bg}}\left(T_e - T_l\right), \\
C_{\mathrm{op}} \frac{dT_{\mathrm{op}}}{dt} &= G_{e\text{-}op}\left(T_e - T_{\mathrm{op}}\right) - \frac{C_{\mathrm{op}}}{\tau_{\mathrm{op}}}\left(T_{\mathrm{op}} - T_l\right),
\end{align}
where $C_e$ and $C_{\mathrm{op}}$ are the effective heat capacities of electrons and optical phonons, $G_{e\text{-}op}$ is the electron-optical-phonon coupling constant, $G_{\mathrm{bg}}$ accounts for background cooling channels (including acoustic-phonon and substrate-mediated pathways), and $\tau_{\mathrm{op}}$ is the lifetime of the optical phonons against decay into the lattice.

It is important to emphasize that the observed excitation photon-energy-dependent resonance window (around  0.42-0.48 eV) emerges despite the purely intraband nature of the probed response. In particular, the Drude conductivity $\sigma_{\mathrm{intra}}(\omega_{\mathrm{THz}}, T_e)$, which governs THz absorption, does not depend explicitly on the pump photon energy ($h\nu_{\mathrm{pump}}$). Instead, within the extended two-temperature model, the excitation photon energy influences the dynamics solely through the source term $S(t)$, which determines the initial energy deposited into the electronic system. Pauli blocking, therefore, affects only the energy-deposition process and the resulting initial non-equilibrium carrier distribution, while the subsequent relaxation dynamics are governed by electron-phonon coupling and optical-phonon decay.

Following optical excitation, photoexcited carriers thermalize via carrier–carrier scattering on sub-100-fs timescales, establishing a hot Fermi-Dirac distribution before the onset of energy relaxation to phonons~\cite{bib7,bib8,bib9}. The absorbed energy per pulse is expressed as $S(t)=\eta(E_{\mathrm{exc}},E_F)$$Fg(t)$, where $F$ is the incident pump fluence and $g(t)$ is the normalized temporal envelope, and $E_{\mathrm{exc}}$ is the excitation photon energy. 

The baseline interband absorption efficiency in the absence of Pauli blocking, denoted $\eta_0(E_{\mathrm{exc}})$ $\approx0.029$, is determined by a single-layer graphene’s universal interband optical conductivity ~\cite{bib34,bib35} and modified by the dielectric environment of the substrate. Fermi filling suppresses interband absorption through Pauli blocking, which we incorporate using a finite-temperature occupation factor

$
\mathcal{P}(E_{\mathrm{exc}},E_F,T)=\tfrac12\left[\tanh\left(\frac{E_{\mathrm{exc}}+2E_F}{4k_BT_e}\right)+\tanh\left(\frac{E_{\mathrm{exc}}-2E_F}{4k_BT_e}\right)\right]$ ~\cite{bib37,bib36}
so that, $\eta(E_{\mathrm{exc}},E_F)=\eta_0(E_{\mathrm{exc}})\mathcal{P}$, consistent with prior theoretical and experimental treatments of photon-energy–\\
dependent absorption in graphene~\cite{bib20,bib22,Dawlaty}.  Here,
 $k_B = 1.380\,649 \times 10^{-23}\ \mathrm{J\,K^{-1}}$ is the Boltzmann constant, and $T$ is the temperature.  
Within this formulation, Pauli blocking determines only the magnitude of the initial electronic heating and thereby specifies how the excitation photon energy enters the model, while the subsequent relaxation dynamics are governed by electron-optical-phonon coupling and optical-phonon decay into the lattice, as described by the coupled energy-balance equations~\cite{bib12,bib14,bib28,bib33}.

For the intraband heating, the absorbance is taken to be proportional to the real part of the intraband optical conductivity within the Drude description, where the Drude weight is given by ~\cite{bib37} (see SI).
\begin{equation}
D(T_e)
=
\frac{2e^2}{\hbar^2}
k_B T_e
\ln\!\left[
2\cosh\!\left(\frac{E_F}{2k_B T_e}\right) 
\right].
\end{equation}

Although the intraband conductivity, which depends on the instantaneous electronic temperature $T_e(t)$, chemical potential, and scattering rate, does not depend explicitly on the excitation photon energy, it determines the initial carrier energy distribution prior to thermalization. However, the pump photon energy critically determines which electronic states are initially populated and how efficiently those carriers can emit optical phonons before thermalization. In graphene, when the pump energy approaches roughly twice the optical-phonon energy ($\sim$0.18 - 0.20 eV), carriers can efficiently emit multiple optical phonons in rapid succession, producing a spike in optical-phonon generation and a transient hot-phonon bottleneck (we explain this in the discussion section). For lower photon energies, this channel is inefficient, while at much higher energies, the larger phase space allows phonon decay to compete effectively, suppressing accumulation. Thus, the observed resonance window around 0.42 - 0.48 eV is a population-driven phonon effect rather than a direct conductivity resonance. Within the extensive two-temperature model, the correct way to incorporate photon-energy dependence is through the source term and the electron-optical-phonon coupling pathway.

This model reduces to the conventional two-temperature description when optical phonons do not accumulate, but it is essential for capturing hot-phonon bottleneck effects when a non-equilibrium optical-phonon population forms. All experimental regimes are described within this single framework by allowing the effective parameters, most notably $\tau_{\mathrm{op}}$, to depend on excitation photon energy.

Having established the unified carrier-phonon energy-balance model, we now show how this framework naturally accounts for the distinct carrier relaxation behaviors observed across the three excitation-energy regimes. All excitation-energy regimes are described within the same extended two-temperature model; the governing equations remain unchanged, while the effective parameters - most notably the optical-phonon lifetime $\tau_{\mathrm{op}}$ - vary with excitation photon energy.

For low excitation energies ($E_{\mathrm{exc}} < 0.42~\mathrm{eV}$), $\tau_{\mathrm{op}} \approx \tau_0$ (typically $\sim 1$ - $2~\mathrm{ps}$ at 295 K) ~\cite{bib11,bib39}, where $\tau_0$ denotes the intrinsic lifetime of graphene optical phonons against anharmonic decay into acoustic phonons and substrate modes. In this regime, $T_{\mathrm{op}} \approx T_l$, and carrier cooling proceeds through fast, irreversible energy transfer (with negligible phonon reabsorption) to the lattice, with a characteristic timescale
\begin{equation}
\tau_{\mathrm{cool}} \approx \frac{C_e}{G_{e\text{-}op} + G_{\mathrm{bg}}},
\end{equation}
corresponding to the standard two-temperature limit.

In contrast, excitation within the narrow resonant window ($0.42 \leq E_{\mathrm{exc}} \leq 0.48~\mathrm{eV}$) leads to a pronounced slowdown of carrier relaxation. Here, optical-phonon generation outpaces decay, resulting in the buildup of a non-equilibrium phonon population whose reabsorption feeds energy back to the electronic system, producing a hot-phonon bottleneck. Within the model, this behavior is captured by a strong enhancement of $\tau_{\mathrm{op}}(E_{\mathrm{exc}})$, causing electrons and optical phonons to equilibrate with each other before cooling to the lattice. The observed long-lived decay corresponds to the slow eigenmode of the coupled system and approaches
\begin{equation}
\tau_{\mathrm{long}} \approx \frac{C_e + C_{\mathrm{op}}}{G_{\mathrm{bg}} + C_{\mathrm{op}}/\tau_{\mathrm{op}}}.
\end{equation}

For higher excitation energies ($E_{\mathrm{exc}} > 0.48~\mathrm{eV}$), the system exits the resonant condition: the electronic phase space is large, multiple optical-phonon emission events are allowed, and phonon decay again competes effectively with generation. As a result, the effective $\tau_{\mathrm{op}}$ decreases toward $\tau_0$, the bottleneck collapses, and fast carrier cooling is restored with
\begin{equation}
\tau_{\mathrm{cool}} \approx \frac{C_e}{G_{e\text{-}op} + G_{\mathrm{bg}}},
\end{equation}
marking a loss of control outside the resonant excitation window. Further details are provided in the supported information.

The pump-probe THz measurements reveal that hot-carrier cooling in graphene is strongly and non-monotonically dependent on the excitation photon energy. While the peak differential transmission scales sublinearly with absorbed photon density across all excitation energies - consistent with hot-carrier heating dynamics - the extracted relaxation times vary dramatically with photon energy, as commonly observed in time-resolved THz photoconductivity studies of graphene~\cite{bib17,bib24,bib26}. At room temperature, optical-phonon emission is generally expected to dominate carrier cooling for excitation energies above the optical-phonon threshold, yielding decay times in the 1 - 5~ps range~\cite{bib12,bib14,bib15}. In stark contrast to this expectation, we observe anomalously long relaxation times, reaching up to $\sim 20$~ps, when the excitation photon energy lies near 0.46~eV (2.7~$\mu$m). To our knowledge, such a pronounced slowdown at room temperature has not been previously reported.

Across a set of monolayer and bilayer graphene samples on silicon carbide and silicon/
silicon-oxide substrates, spanning a range of Fermi energies from 0.06 to 0.39 eV, the relaxation time displays a pronounced non-monotonic dependence on the excitation photon energy. Specifically, within a narrow spectral window from 0.42 to 0.48 eV ($\approx$70 meV wide), the relaxation time increases by nearly an order of magnitude relative to the values observed at both lower and higher photon energies. Outside this window, the decay times remain within the expected few-picosecond range, consistent with earlier broadband pump-probe studies~\cite{bib15}. The reproducibility of this behavior across samples indicates that it reflects an intrinsic, energy-selective cooling mechanism rather than extrinsic disorder or substrate effects~\cite{bib13}.

To elucidate the origin of this behavior, it is instructive to distinguish three excitation-energy regimes. For excitation energies below $\sim 0.42~\text{eV}$, (Case I) (Fig.4a),  carriers possess only a small excess energy above the Fermi level following rapid electron-electron thermalization. In this regime, the initial carrier energy relative to the Fermi level,
$\Delta\varepsilon = \varepsilon - E_F$, satisfies
\begin{equation}
\hbar\omega_{\mathrm{op}} < \Delta\varepsilon < 2\hbar\omega_{\mathrm{op}},
\end{equation}
allowing the emission of at most a single optical phonon. Once this emission occurs, the carrier energy falls below $\hbar\omega_{\mathrm{op}}$, and further optical-phonon emission is energetically forbidden. Consequently, relaxation proceeds predominantly via low-energy acoustic phonons or weaker scattering channels. Phase-space restrictions play a minor role, as most conduction-band states above the Fermi level remain available. The few optical phonons generated decay efficiently into acoustic modes and the substrate without significant accumulation, acting as an effective energy sink rather than a bottleneck. As a result, carrier cooling remains relatively fast, with relaxation times in the few-picosecond range, limited primarily by the low excitation energy rather than by hot-phonon reabsorption.

A qualitatively different behavior emerges when the excitation photon energy lies within the narrow window between 0.42 and 0.48~eV (Case~II) (Fig.4b, right side of the cartoon). In this regime, the pump photon energy approaches twice the optical-phonon energy and, for intermediate-doped samples, closely matches the Pauli-blocking threshold $2E_F$~\cite{bib21,bib22}. Under these resonant conditions, optical-phonon generation is particularly effective, while competing relaxation pathways are partially suppressed or redistributed. As carriers retain sufficient energy after phonon emission to access higher-lying electronic states, the balance between phonon generation and decay is disrupted, leading to the buildup of a non-equilibrium optical-phonon population. Reabsorption of these hot phonons feeds energy back into the electronic system, producing a pronounced hot-phonon bottleneck and a three-to four-fold increase in the carrier relaxation time, consistent with theoretical predictions of phonon-mediated cooling suppression~\cite{bib12,bib28}.

Importantly, this anomalous slowdown is observed across all samples, regardless of Fermi energy, indicating that Pauli blocking alone cannot account for the effect. In intermediate-doped samples, partial Pauli blocking suppresses interband excitation and favors intraband heating~\cite{bib22,Tielrooij}, which funnels relaxation into a narrower set of phonon-mediated channels and enhances phonon accumulation. However, even in low-doped samples where interband transitions remain fully allowed, and in highly doped samples where excitation is entirely intraband, the long-lived dynamics persist within the same excitation window. The common feature across all cases is the simultaneous presence of efficient optical-phonon generation, sufficient electronic phase space for reabsorption, and suppression or redistribution of competing cooling pathways, leading to a population-driven hot-phonon bottleneck ~\cite{bib15,bib28,bib33}.

\begin{figure}[!htbp]
\centering
\includegraphics[scale=0.66]{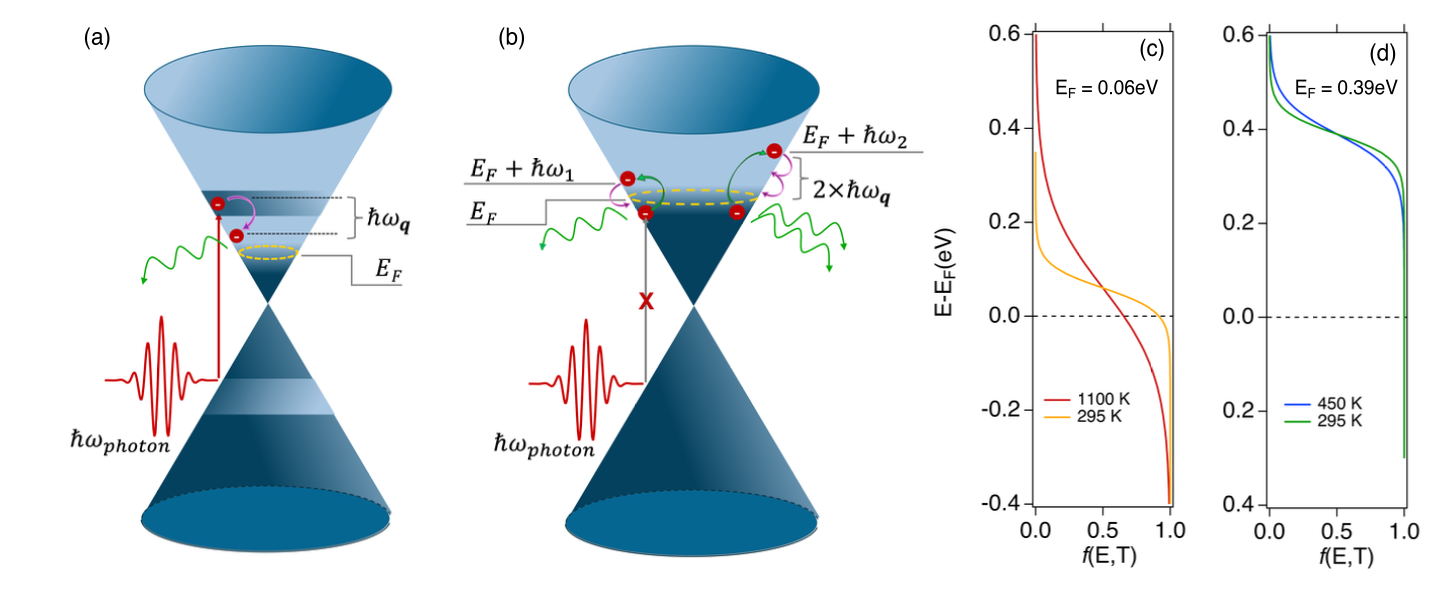}
\caption{Doping- and photon-energy-dependent carrier relaxation pathways in graphene.  (a) Low-doping regime ($E_F = 0.06$ eV), where Pauli blocking is absent and interband absorption is allowed. Photoexcitation with photon energies below the resonant window ($\hbar \omega < 0.4$ eV; red vertical arrow) generates electron-hole pairs. Carrier relaxation proceeds via emission of at most one optical phonon ($\hbar \omega_{\mathrm{OP}} \approx 0.2$ eV; purple curve) followed by decay into acoustic phonons (green wavy line), resulting in fast cooling on a few-picosecond timescale. b) High-doping regime ($E_F = 0.39$ eV), where interband transitions are Pauli blocked for $\hbar \omega < 0.4$ eV and photoexcitation leads to intraband carrier heating (green curves). Excitation with photon energy $\hbar \omega_1$ to energies $E_F + \hbar \omega_1$ results in optical-phonon emission followed by anharmonic decay into acoustic modes, yielding fast cooling on a few-picosecond timescale. In contrast, excitation with photon energy within the resonance window $\hbar \omega_2$, which promotes carriers to energies near twice the optical-phonon energy, leads to rapid optical-phonon generation (two OP) and accumulation due to relatively small phase space. When optical-phonon emission exceeds phonon decay, a hot-phonon bottleneck forms, enabling phonon reabsorption and confining carrier energies near $E_F + \hbar \omega_{\mathrm{OP}}$. Tuning the excitation photon energy to $\sim 0.46$ eV activates this recycling regime, extending carrier relaxation times to $\sim 15$-20 ps. (c) Calculated Fermi-Dirac distributions for the low-doping case at the lattice temperature (295 K, yellow curve) and at an elevated electronic temperature ($\sim 1100$ K, red curve), assuming absorption of $\sim 0.023$ of a 3~$\mu$J cm$^{-2}$ excitation fluence. (d) Calculated Fermi-Dirac distributions for the high-doping case at 295 K (green curve) and at an elevated electronic temperature ($\sim 450$ K, blue curve).}\label{Fuencedependence}
\end{figure}

At higher excitation energies above $\sim 0.48$~eV (Case~III) (Fig.4b left side of the cartoon), fast carrier cooling is restored across all samples, and the photoexcited carriers populate a broad energy distribution extending well above the Fermi level. In this regime, optical-phonon emission ~\cite{bib12,bib27} does not confine carriers near $E_F$, enabling multiple successive emission events into a large manifold of available electronic states. The expanded phase space suppresses phonon reabsorption~\cite{bib11,bib25}, and allows optical-phonon decay to effectively compete with generation, preventing sustained phonon accumulation. As a result, the hot-phonon bottleneck collapses and carrier cooling returns to the intrinsic few-picosecond timescale.

For clarity, figures 4(c) and 4(d) show the calculated Fermi–Dirac distributions in the low- and high-doping regimes, respectively. For the low-doping case (Fig. 4c), the equilibrium carrier distribution at the lattice temperature of 295~K (yellow curve) evolves to a hot-electron distribution corresponding to an electronic temperature of approximately 1100~K (red curve), assuming absorption of $\sim$0.023 of an excitation fluence of 3~$\mu$J~cm$^{-2}$. In the high-doping case (Fig. 4d), the equilibrium distribution at 295~K (green curve) increases to a more moderate electronic temperature of approximately 450~K (blue curve) under the same excitation conditions.

Although interband excitation in lightly doped graphene creates electron-hole pairs, the observed relaxation dynamics are governed by cooling of the hot electronic distribution via optical-phonon emission and reabsorption, while interband recombination plays a negligible role on the picosecond timescale.
Taken together, these results demonstrate that resonant optical-phonon excitation provides a direct and energy-selective means to control hot-carrier cooling in graphene. By tuning the excitation photon energy, it is possible to reversibly switch between fast and strongly suppressed cooling regimes at room temperature, without modifying material composition or carrier density. This energy-selective control highlights the central role of optical-phonon dynamics in non-equilibrium energy flow and opens new opportunities for engineering ultrafast carrier lifetimes in graphene-based optoelectronic and terahertz devices~\cite{bib31,bib32,bib3}.

In conclusion, our measurements reveal that hot-carrier cooling in graphene can be switched between fast and strongly suppressed regimes by selectively tuning the excitation photon energy. While carrier relaxation remains in the few-picosecond range over a broad span of excitation energies and fluences, a pronounced and reproducible slowdown emerges only within a narrow window centered near 0.45 eV, where optical-phonon accumulation and reabsorption dominate the energy-dissipation pathway. Crucially, the governing carrier-phonon interactions remain unchanged; instead, excitation energy controls the effectiveness of optical-phonon decay, enabling reversible access to a hot-phonon-bottlenecked regime at room temperature. These findings establish excitation-energy-selective control of non-equilibrium carrier dynamics in graphene and highlight optical phonons as a powerful handle for engineering ultrafast energy flow in two-dimensional materials and related optoelectronic platforms.

\section*{Acknowledgements}

I. C. acknowledges support from start-up funds provided by Texas Tech University. Research at the U.S. Naval Research Laboratory was supported by the Office of Naval Research
(ONR).


\section*{\Huge{Supporting information}}


\section*{S1. Extended carrier-phonon energy-balance model}

The ultrafast carrier dynamics in graphene are described using an extended two-temperature model that explicitly accounts for a nonequilibrium population of strongly coupled optical phonons ~\cite{ref1,ref2,ref3,ref4,ref5,ref6}. The model consists of three subsystems: (i) the electronic system, characterized by an effective temperature $T_e(t)$; (ii) an optical-phonon bath with effective temperature $T_{\mathrm{op}}(t)$; and (iii) the remaining lattice degrees of freedom (acoustic phonons and substrate), treated as a thermal reservoir at fixed temperature $T_l$ ~\cite{ref2,ref7}. 

The coupled energy-balance equations are
\begin{equation}
C_e \frac{dT_e}{dt}
=
S(t)
-
G_{e\text{-}op}\left(T_e - T_{\mathrm{op}}\right)
-
G_{\mathrm{bg}}\left(T_e - T_l\right),
\tag{S1}
\end{equation}

\begin{equation}
C_{\mathrm{op}} \frac{dT_{\mathrm{op}}}{dt}
=
G_{e\text{-}op}\left(T_e - T_{\mathrm{op}}\right)
-
\frac{C_{\mathrm{op}}}{\tau_{\mathrm{op}}}
\left(T_{\mathrm{op}} - T_l\right),
\tag{S2}
\end{equation}

where $C_e$ and $C_{\mathrm{op}}$ are the effective heat capacities of the electronic and optical-phonon subsystems, respectively ~\cite{ref9}; $G_{e\text{-}op}$ is the electron--optical-phonon coupling constant ~\cite{ref5,ref10}; $G_{\mathrm{bg}}$ accounts for background cooling channels, including coupling to acoustic phonons and the substrate ~\cite{ref11}; and $\tau_{\mathrm{op}}$ is the lifetime of optical phonons against decay into the lattice~\cite{ref13,ref14,ref15}.

The pump excitation is introduced through the source term $S(t)$, which represents the rate of energy deposited into the electronic system~\cite{ref1,ref16}. The magnitude of $S(t)$ depends on the excitation photon energy, pump fluence, and Fermi level through Pauli blocking of interband transitions~\cite{ref17,ref18,ref19}. Within this framework, Pauli blocking affects only the initial energy deposition via $S(t)$ and does not modify the subsequent relaxation equations consistent with previous theoretical treatments~\cite{ref5,ref18}.

\subsection*{S2. Linearized form and matrix representation}

For times exceeding the initial electronic thermalization ($\gtrsim 100~\mathrm{fs}$), the carrier distribution can be described by a hot Fermi–Dirac distribution characterized by a single electronic temperature \(T_e(t)\)~\cite{ref1,ref20}. 

Linearizing Eqs.~(S1)--(S2) around the lattice temperature by defining

\begin{equation}
\Delta T_e(t) = T_e(t) - T_l,
\qquad
\Delta T_{\mathrm{op}}(t) = T_{\mathrm{op}}(t) - T_l.
\end{equation}
the temperature deviations yields a linear time-invariant system \(T_l\) ~\cite{ref6,ref21}.
Equations~(S1)--(S2) then take the linear time-invariant form
\begin{equation}
\frac{d}{dt}
\begin{pmatrix}
\Delta T_e \\
\Delta T_{\mathrm{op}}
\end{pmatrix}
=
\mathbf{A}
\begin{pmatrix}
\Delta T_e \\
\Delta T_{\mathrm{op}}
\end{pmatrix}
+
\mathbf{B}\,u(t),
\tag{S3}
\end{equation}

with
\begin{equation}
\mathbf{A}
=
\begin{pmatrix}
-\dfrac{G_{e\text{-}op}+G_{\mathrm{bg}}}{C_e}
&
\dfrac{G_{e\text{-}op}}{C_e}
\\[6pt]
\dfrac{G_{e\text{-}op}}{C_{\mathrm{op}}}
&
-\left(
\dfrac{G_{e\text{-}op}}{C_{\mathrm{op}}}
+
\dfrac{1}{\tau_{\mathrm{op}}}
\right)
\end{pmatrix},
\qquad
\mathbf{B}
=
\begin{pmatrix}
\dfrac{1}{C_e} \\
0
\end{pmatrix}.
\tag{S4}
\end{equation}

as commonly used in coupled electron--phonon thermal models~\cite{ref6,ref21}.

Here, $u(t)$ describes the temporal profile of the pump-induced energy deposition.

\subsection*{S3. Exact analytical solution}

The system described by Eq.~(S3) admits an exact analytical solution via the variation-of-constants (Duhamel) formalism for linear time-invariant systems~\cite{ref23,ref24}. For an initial condition \(\mathbf{z}(t_0)\), the solution is

\[
\mathbf{z}(t_0)
=
\begin{pmatrix}
\Delta T_e(t_0) \\
\Delta T_{\mathrm{op}}(t_0)
\end{pmatrix},
\]
the solution is
\begin{equation}
\mathbf{z}(t)
=
e^{\mathbf{A}(t-t_0)}\mathbf{z}(t_0)
+
\int_{t_0}^{t}
e^{\mathbf{A}(t-s)} \mathbf{B}\,u(s)\,ds.
\tag{S5}
\end{equation}

In pump--probe experiments, the source term vanishes for $t \ge t_0$, and the long-time dynamics are governed entirely by the homogeneous solution ~\cite{ref21}.
\begin{equation}
\mathbf{z}(t)
=
e^{\mathbf{A}(t-t_0)}\mathbf{z}(t_0).
\tag{S6}
\end{equation}

\subsection*{S4. Eigenmodes and relaxation times}

Because $\mathbf{A}$ is a $2\times2$ matrix, the homogeneous solution can be expressed as a sum of two exponential modes ~\cite{ref23},
\begin{equation}
\Delta T_e(t)
=
A_1 e^{-t/\tau_1}
+
A_2 e^{-t/\tau_2},
\tag{S7}
\end{equation}
where $\tau_{1,2} = -1/\lambda_{1,2}$ are determined by the eigenvalues $\lambda_{1,2}$ of $\mathbf{A}$.

The eigenvalues are
\begin{equation}
\lambda_{\pm}
=
-\frac{1}{2}
\left[
\frac{G_{e\text{-}op}+G_{\mathrm{bg}}}{C_e}
+
\frac{G_{e\text{-}op}}{C_{\mathrm{op}}}
+
\frac{1}{\tau_{\mathrm{op}}}
\right]
\pm
\frac{1}{2}
\sqrt{
\left(
\frac{G_{e\text{-}op}+G_{\mathrm{bg}}}{C_e}
-
\frac{G_{e\text{-}op}}{C_{\mathrm{op}}}
-
\frac{1}{\tau_{\mathrm{op}}}
\right)^2
+
\frac{4G_{e\text{-}op}^2}{C_e C_{\mathrm{op}}}
}.
\tag{S8}
\end{equation}

consistent with prior treatments of hot-phonon bottlenecks in graphene~\cite{ref5,ref6,ref13}. The experimentally observed relaxation time corresponds to the slowest decay constant,
$\tau_{\mathrm{long}} = \max(\tau_1,\tau_2)$.

\subsection*{S5. Limiting cases}

\paragraph{Fast optical-phonon decay.}
In the limit of rapid optical-phonon decay
($\tau_{\mathrm{op}} \ll C_{\mathrm{op}}/G_{e\text{-}op}$),
optical phonons remain in equilibrium with the lattice
($T_{\mathrm{op}} \approx T_l$), and the model reduces to a single-temperature description governed by background cooling~\cite{ref2}:
\begin{equation}
\boxed{\tau_{\mathrm{cool}}
\approx
\frac{C_e}{G_{e\text{-}op}+G_{\mathrm{bg}}}
\tag{S9}
}
\end{equation}

\paragraph{Hot-phonon bottleneck regime.}
In the opposite limit of suppressed effective optical-phonon decay
($\tau_{\mathrm{op}} \gg C_{\mathrm{op}}/G_{e\text{-}op}$),
electrons and optical phonons equilibrate prior to cooling to the lattice ($T_e \approx T_{\mathrm{op}}$), yielding the hot-phonon bottleneck regime ~\cite{ref5,ref13,ref14,ref15}
\begin{equation}
\boxed{\tau_{\mathrm{long}}
\approx
\frac{C_e + C_{\mathrm{op}}}{G_{\mathrm{bg}} + C_{\mathrm{op}}/\tau_{\mathrm{op}}}.
\tag{S10}
}
\end{equation}

\begin{itemize}
    \item \textbf{Fast OP decay:} Optical phonons relax quickly to the lattice. Electron cooling is dominated by background channels (\(G_{\mathrm{bg}}\)).
    \item \textbf{Hot-phonon bottleneck:} Electrons and optical phonons equilibrate first; cooling is limited by the slow decay of optical phonons (\(1/\tau_{\mathrm{op}}\)) to the lattice.
\end{itemize}

\section*{Physical meaning}

\begin{itemize}
\item $G_{e\text{-}op}$ controls how fast electrons and optical phonons equilibrate.
\item $\tau_{\mathrm{op}}$ controls how fast energy leaves the optical-phonon bath.
\item A long relaxation time arises when energy is trapped in optical phonons,
not when electron--phonon coupling is weak.
\end{itemize}

In other words, in the bottleneck regime, increasing $G_{e\text{-}op}$ accelerates equilibration
between electrons and optical phonons but does not enhance energy dissipation
to the lattice.
Instead, the two subsystems behave as a single composite reservoir with heat
capacity $C_e+C_{\mathrm{op}}$, whose cooling rate is limited by optical-phonon
anharmonic decay ~\cite{ref25} and background coupling.
As a result, stronger electron--phonon coupling can increase the effective
thermal inertia of the relaxing system and thereby lengthen the observed decay
time.

\subsection*{S6. Source Term}
We model the pump energy deposition self-consistently via
\begin{equation}
S(t)=A(\omega,T_e(t))\,I(t),
\end{equation}
where $I(t)$ is a 100-fs-FWHM Gaussian intensity envelope normalized to the
measured fluence, and $A(\omega,T_e)$ includes interband Pauli blocking and
intraband absorption  ~\cite{ref26}.

The source term is written as
\begin{equation}
S(t)=A(\omega,T_e(t))\,I(t).
\tag{S11}
\end{equation}

The total absorbance is decomposed as
\begin{equation}
A(\omega,T_e)=A_{\mathrm{inter}}(\omega,T_e)+A_{\mathrm{intra}}(\omega,T_e).
\tag{S12}
\end{equation}

\subsection*{(i) Interband Absorbance}

The interband absorbance with an instantaneous Pauli-blocking factor is
\begin{equation}
A_{\mathrm{inter}}(\omega,T_e)
=
A_0\,\mathcal{P}_{\mathrm{inter}}(\omega,\mu,T_e),
\qquad
A_0\simeq 0.023.
\tag{S13}
\end{equation}

The Pauli-blocking factor is given by
\begin{equation}
\mathcal{P}_{\mathrm{inter}}(\omega,\mu,T_e)
=
\frac{1}{2}
\left[
\tanh\!\left(\frac{\hbar\omega+2E_F}{4k_B T_e}\right)
+
\tanh\!\left(\frac{\hbar\omega-2E_F}{4k_B T_e}\right)
\right].
\tag{S14}
\end{equation}

\subsection*{(ii) Intraband Absorbance}

The intraband absorbance is taken to be proportional to the real part of the
intraband optical conductivity,
\begin{equation}
%
A_{\mathrm{intra}}(\omega,T_e)
=
\frac{4Z_0}{(1+n_s)^2}\,
\Re\!\left[\sigma_{\mathrm{intra}}(\omega,T_e)\right],
\tag{S15}
\end{equation}

Within a Drude description,
\begin{equation}
\Re\!\left[\sigma_{\mathrm{intra}}(\omega,T_e)\right]
=
\frac{D(T_e)}{\pi}\,
\frac{\Gamma}{\omega^2+\Gamma^2},
\tag{S16}
\end{equation}
where the Drude weight is
\begin{equation}
D(T_e)
=
\frac{2e^2}{\hbar^2}
k_B T_e
\ln\!\left[
2\cosh\!\left(\frac{E_F}{2k_B T_e}\right)
\right].
\tag{S17}
\end{equation}

\subsection*{Parameter values and fitting procedure}

Typical starting values used in fitting were
$C_e \sim 10^{-6}$--$10^{-5}~\mathrm{J\,m^{-2}\,K^{-1}}$,
$C_{\mathrm{op}} \sim 10^{-5}~\mathrm{J\,m^{-2}\,K^{-1}}$,
$G_{e\text{-}op} \sim 10^{7}$--$10^{8}~\mathrm{W\,m^{-2}\,K^{-1}}$,
$G_{\mathrm{bg}} \sim 10^{6}$--$10^{7}~\mathrm{W\,m^{-2}\,K^{-1}}$,
and $\tau_{\mathrm{op}} \sim 0.5$--$10~\mathrm{ps}$ ~\cite{ref9,ref10,ref13,ref27}.
Heat capacities and coupling constants were treated as global parameters across excitation energies, while $\tau_{\mathrm{op}}$ was allowed to vary with excitation photon energy consistent with prior hot-phonon analyses ~\cite{ref3}. Fits were performed globally using the slow eigenmode extracted from Eq.~(S8).

\subsection*{Experimetal-Samples preparation}

Experiments were conducted on monolayer and bilayer graphene samples fabricated using two complementary growth approaches. Monolayer graphene was synthesized by chemical vapor deposition (CVD) and subsequently transferred onto SiO$_2$/Si substrates with a 280 nm oxide layer. Bilayer graphene consisted of quasi-freestanding bilayer graphene grown epitaxially on semi-insulating 6H-SiC(0001) substrates via thermal decomposition. Hydrogen intercalation was then employed to decouple the graphene layers from the SiC substrate.

The epitaxial graphene was grown in an Axitron/Epigress VP508 horizontal hot wall reactor at 1540~$^\circ$C in 100~mbar Ar for 12~minutes. Prior to graphene growth, the SiC substrate was heated under high-purity hydrogen to remove substrate polishing damage. After growth, hydrogen intercalation was performed \textit{in situ} at 1050~$^\circ$C in 900~mbar hydrogen for 1~hour.

\begin{figure}[!htbp]
\centering
\includegraphics[scale=0.62]{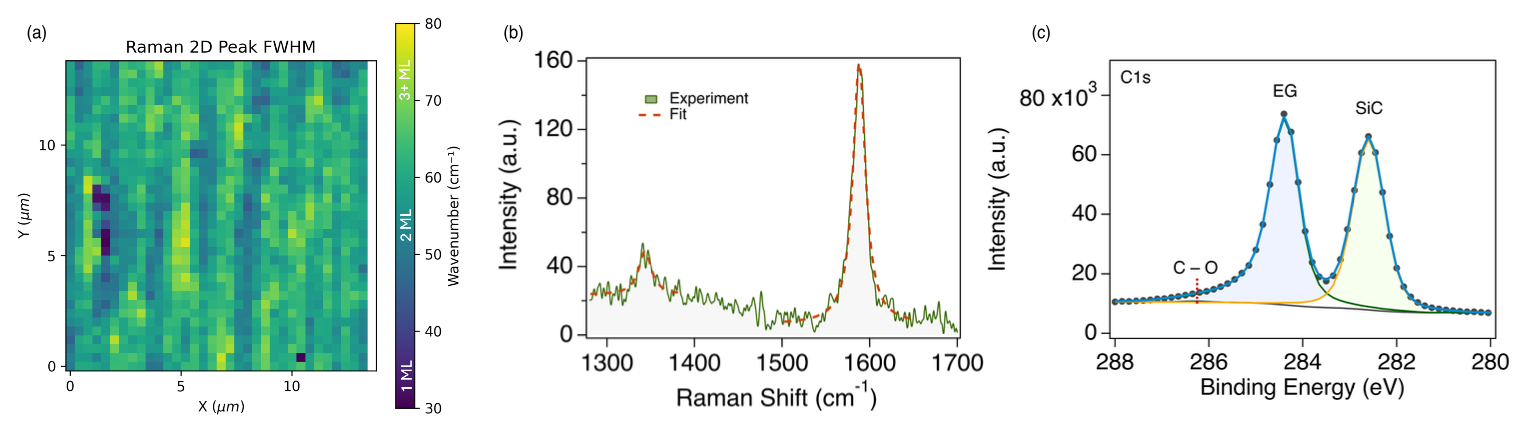}
\caption{Spectroscopic characterization of epitaxial graphene on 6H-SiC and CVD graphene on SiO$_2$/Si.
(a) Raman mapping based on the 2D-band linewidth of epitaxial graphene indicating that the film is predominantly bilayer graphene across the mapped region. (b) Raman spectrum of CVD monolayer graphene on SiO$_2$/Si.(c) XPS C 1s core level spectrum of epitaxial graphene.
}\label{Photondependence}.
\end{figure}

A Raman map of the 2D peak full width at half maximum (FWHM) (Fig.~\ref{Photondependence}a) reveals that the hydrogen-intercalated epitaxial graphene consists predominantly of bilayer graphene on the terraces, while thicker graphene regions are observed near the step edges. The average 2D peak FWHM is 59~cm$^{-1}$~\cite{Lee}. This value is consistent with expectations for epitaxial graphene grown on SiC and subsequently hydrogen-intercalated, where hydrogen atoms decouple the interfacial buffer layer from the SiC substrate and convert it into a second graphene layer~\cite{Daniels}.
Raman mapping was performed using a Thermo DXRxi Raman microscope equipped with a 532~nm excitation laser (9.6~mW) and a 100$\times$ objective. The FWHM of the 2D peak at each pixel was extracted by fitting the spectral region around the 2D peak with a Lorentzian function on a linear background. Electrical transport properties were characterized using Van der Pauw Hall measurements performed at room temperature with a home-built system employing a 2060~G magnetic field and four contacts placed at the perimeter of the $8 \times 8~\mathrm{mm}^2$ sample.

X-ray photoelectron spectroscopy (XPS) was employed to characterize the epitaxial graphene and its interaction with the 6H-SiC substrate after hydrogen intercalation. The C~1$s$ core level spectrum (Fig.~\ref{Photondependence}b) exhibits peaks corresponding to sp$^2$ carbon in graphene (labeled “EG”) at 284.4~eV and Si-C bonds in the substrate (labeled “SiC”) at 282.6~eV. The large separation between the “SiC” and “EG” peaks is characteristic of hydrogen-intercalated quasi-freestanding epitaxial graphene, where a shift in the graphene peak to lower binding energy results from p-type doping, and the shift in the “SiC” peak is due to surface band bending upon intercalation~\cite{Mammadov,Riedl}. Additionally, the absence of the interfacial buffer-layer-related components “S1” and “S2” from the C~1$s$ spectra is indicative of decoupling of the graphene from the SiC substrate~\cite{Mammadov,Riedl}. Such interfacial passivation and decoupling reduces substrate-induced scattering and enables the investigation of the intrinsic carrier cooling dynamics of bilayer graphene. 

X-ray photoelectron spectroscopy (XPS) data was collected using a Thermo Scientific K-Alpha instrument with a monochromatic Al-K$\alpha$ source (1486.68~eV) and a 400~$\mu$m x-ray spot size. A flood gun was used for charge compensation. The C~1$s$ core-level spectra were collected using a 50~eV pass energy and 0.15~eV step size. Peak fitting was performed using a Shirley background and convolutions of Gaussian-Lorentzian line shapes. An asymmetric peak shape was used to fit the “graphene” peak in the C~1$s$ spectra. The spectrum was referenced to the position of the C~1$s$ “SiC” peak at 282.6~eV~\cite{Kotsakidis}.

Monolayer graphene samples were prepared by chemical vapor deposition (CVD) and transferred onto SiO$_2$/Si substrates with a $280$~nm oxide layer. For one monolayer sample, the carrier density was determined by Hall measurements to be $2.2 \times 10^{11}\,\mathrm{cm^{-2}}$, corresponding to a Fermi energy of approximately $0.06$~eV. For a second monolayer sample, the Fermi energy was estimated to be approximately $0.18$~eV from Raman spectroscopy Fig.~\ref{Photondependence}c). All samples were measured under ambient conditions at room temperature.

\subsection*{Experimetal-ultrafast optical-pump THz-probe spectroscopy}

Ultrafast measurements were performed using a Ti:sapphire regenerative amplifier operating at a repetition rate of $1$~kHz and delivering pulses with energies of approximately $7$~mJ, pulse durations below $100$~fs, and a central wavelength of $800$~nm. The output of the amplifier was divided into four beams: two for terahertz (THz) generation and detection, one for optical or mid-infrared (MIR) excitation, and one for electro-optic sampling.

Tunable MIR pump pulses were generated using an optical parametric amplifier, providing photon energies between $0.22$ and $0.73$~eV. The pump beam was focused onto the sample to a spot diameter of approximately $1.8$~mm, defined by a metallic aperture placed directly in front of the sample. Unless otherwise stated, measurements were performed at pump fluences of $3$ - $4~\mu\mathrm{J\,cm^{-2}}$.

Broadband THz pulses were generated by optical rectification in a $1$~mm-thick ZnTe (110) crystal. The resulting quasi-single-cycle THz radiation had a bandwidth of approximately $2.3$~THz, centered near $1.2$~THz. The THz beam was focused onto the graphene samples at normal incidence using an off-axis parabolic mirror, producing a spot diameter of approximately $0.7$~mm, as determined by the knife-edge method.

The transmitted THz radiation was collected and refocused onto a second $1$~mm-thick ZnTe (110) crystal for electro-optic sampling. A portion of the $800$~nm beam served as the sampling pulse, and the temporal profile of the THz electric field was recorded by scanning a mechanical delay line. Representative THz waveforms and Fourier spectra, along with the principle of the experimental set up are shown in Figure 1 (main text).

To improve the signal-to-noise ratio, the THz beam was mechanically chopped at $500$~Hz, and the detected signal was recorded using a lock-in amplifier referenced to the chopping frequency. The pump beam was modulated independently using a second mechanical chopper.
Pump-induced changes in THz transmission were measured by varying the delay between the optical pump and the peak of the THz probe pulse. Time-domain THz signals were windowed prior to Fourier transformation to suppress etalon reflections. Unless otherwise stated, pump-probe signals were recorded at a delay of approximately $400$~fs after time zero, exceeding the carrier thermalization time and ensuring that the electronic system had reached a quasi-equilibrium Fermi-Dirac distribution.
Control experiments were performed on bare SiC, sapphire, and SiO$_2$/Si substrates under identical excitation conditions. No pump-induced THz signal was detected in these measurements, confirming that the observed dynamics originate from the graphene layers.

\bibliographystyle{unsrt}

\end{document}